\documentclass[a4paper,12pt]{article}
\usepackage{amsfonts}
\usepackage[utf8]{inputenc}
\usepackage{latexsym,mathrsfs}
\usepackage{amssymb,amsmath,amsthm}
\usepackage{geometry}
\usepackage{graphicx}
\usepackage{multirow}
\usepackage{caption}
\usepackage{subcaption}
\usepackage{mathtools}
\usepackage{pb-diagram}
\usepackage{amscd}

\begin{document}

\date{November 2023}
\title{\textbf{Tsarev Generalized Hodograph Method and Isomonodromic
Solutions of Integrable Dispersive Systems}}
\author{Zakhar V. Makridin, Maxim V. Pavlov}
\date{}
\maketitle

\begin{abstract}
General and particular solutions of the so called semi-Hamiltonian
hydrodynamic type systems can be obtained by the Tsarev Generalized
Hodograph Method. Here we show that a natural extension of this approach
applied to dispersive integrable systems is determined by isomonodromic
deformations.
\end{abstract}

%\begin{keywords}
%Внутренние волны, аттракторы, нагнетаемая мощность.
%\end{keywords}

\vspace{-5mm}

\begin{center}
Institute for Advanced Study\\[0pt]
Shenzhen University\\[0pt]
Shenzhen 518060, Guangdong, P.R. China \\[0pt]
\ \\[0pt]
Department of Mathematics\\[0pt]
Ningbo University\\[0pt]
Ningbo 315211, Zhejiang, P.R. China \\[0pt]
\ \\[0pt]
e-mails: \\[1ex]
\texttt{makridinzv@gmail.com}\\[0pt]
\texttt{mpavlov@itp.ac.ru}
\end{center}

\bigskip


\bigskip


\newpage

\bigskip

\centerline{\it Dedicated to 65th anniversary of our friend Professor Sergey
Tsarev}

%\centerline{\it Dedicated to 65th anniversary of our friend Professor Sergey Tsarev}

\section{Introduction}

The theory of semi-Hamiltonian hydrodynamic type systems (i.e. integrable by
the Tsarev Generalized Hodograph Method) was established in 1985~\cite%
{Ts1985,Ts1991}. Shortly speaking, such a hydrodynamic type system%
\begin{equation}
u_{i,t}=v_{j}^{i}(\mathbf{u})u_{j,x},\text{ \ }i,j=1,2,...,N  \label{hydro}
\end{equation}%
possesses a general solution presented in the algebraic form%
\begin{equation}
x\delta _{k}^{i}+tv_{k}^{i}(\mathbf{u})=w_{k}^{i}(\mathbf{u}),  \label{tghm}
\end{equation}%
where $w_{k}^{i}(\mathbf{u})$ is a velocity matrix (parameterized by $N$
arbitrary functions of a single variable) of the commuting hydrodynamic type
system%
\begin{equation*}
u_{i,\tau }=w_{j}^{i}(\mathbf{u})u_{j,x},\text{ \ }i,j=1,2,...,N.
\end{equation*}%
Here all field variables $u_{k}$ depend on $x,t$ and $\tau $ simultaneously (%
$x$ and $t$ can be interpreted as spatial and temporal variables, while $%
\tau $ is a group parameter).

Most of well-known integrable dispersive systems possess the so called
dispersionless limit and multi-phase solutions. In this paper we restrict
our consideration to this type of integrable dispersive systems only. In a
dispersionless limit they reduce themselves into semi-Hamiltonian
hydrodynamic type systems.
%Also application of the Whitham averaging method to multi-phase solutions yields again semi-Hamiltonian hydrodynamic type systems.

In 1994 B.A. Dubrovin wrote a seminal paper devoted to the relationship
between WDVV associativity equations in Topological Field Theory and Egorov
Hamiltonian hydrodynamic type systems \cite{Dub}. One of his interests was
concentrated on reconstruction of higher order terms (dispersive
corrections) to these systems with preservation of integrability properties
in sense of the Inverse Scattering Transform. Plenty of papers were devoted
to the so called Topological Recursion (higher genus corrections) in past 30
years. Especially advantage was obtained in the Dubrovin--Frobenius
manifolds, associated with isomonodromic deformations, and with the concept
of tau-function naturally appeared in this context \cite{Krichever1, Krichever2}.

In this paper we continue investigation of the Dubrovin--Frobenius
manifolds, but not as a geometrical object. Here we deal with the formal
procedure: how to reconstruct higher order terms (dispersive corrections) to
solutions of Egorov Hamiltonian hydrodynamic type systems appeared in the
Topological Field Theory in the framework suggested by B.A. Dubrovin in 1994.

Thus Our \textbf{Target} is to suggest extension of the Tsarev Generalized
Hodograph Method to integrable (by the Inverse Scattering Transform Method)
dispersive systems, which have a dispersionless limit.

A.B. Shabat and A.P. Veselov~\cite{VeSha1993} investigated solutions of the
Korteweg--de Vries equation (KdV) by analysis of closure conditions for
the corresponding dressing chain in 1993. Their result was following: the space
of finite-gap solutions (also known as multi-phase solutions or
algebro-geometric solutions) of the KdV equation is parameterized by $2N+1$
arbitrary constants ($N$ is a genus of corresponding hyperelliptic Riemann
surface), while the space of isomonodromic deformations is parameterized by $%
2N+2$ arbitrary constants. This means that the closure condition of the KdV
dressing chain contains the exceptional arbitrary parameter $\alpha $ such
that: the degenerated case $\alpha =0$ contains finite-gap solutions only,
while the generic case $\alpha \neq 0$ leads to isomonodromic deformations.
So finite-gap solutions of the KdV equation can be obtained by a limit
from isomonodromic solutions of the KdV equation.

However \textit{algebro-geometric solutions disappear in a dispersionless
limit}. This means that they (i.e. finite-gap solutions) cannot be
considered as a candidate to be an extension for the Tsarev algebraic
formula (\ref{tghm}).

Our \textbf{Statement} is: \textit{isomonodromic deformations of integrable
dispersive systems (i.e. determined by Lax representations) present a
natural extension of the Tsarev algebraic formula} (\ref{tghm}) \textit{to a
ordinary differential equations}:%
\begin{equation}
x\delta _{k}^{i}+tv_{k}^{i}(\mathbf{u,u}_{x}\mathbf{,u}_{xx},...)=w_{k}^{i}(%
\mathbf{u,u}_{x}\mathbf{,u}_{xx},...).  \label{etghm}
\end{equation}

This means, that the above system of ordinary differential equations under
the rescaling $\partial _{x}\rightarrow \epsilon \partial _{x},\partial
_{t}\rightarrow \epsilon \partial _{t}$ and a formal limit $\epsilon
\rightarrow 0$ implies the Tsarev algebraic formula (\ref{tghm}).

For one-component case ($i=1$) equations of the form~\eqref{etghm} possess
an important class of solutions first appeared in the context of the Cauchy
problem for the KdV equation in small-dispersion limit~\cite{GP1973}. It
turns out, that in the vicinity of breaking point the solution of KdV
equation
\begin{equation*}
u_t+6uu_x+u_{xxx}=0
\end{equation*}
satisfies the fourth-order ODE of the Painlev\'{e} I hierarchy~\cite%
{Sul1994,KudSul1996,Dubr2006,ClaGra2009}:
\begin{equation*}
x-6tu=-u^3-\frac{1}{2}u_x^2-uu_{xx}-\frac{1}{10}u_{xxxx}.
\end{equation*}
After breaking point the dynamics is more complicated: it includes developing
of oscillation zone, which can be described by multiscale asymptotics~\cite%
{GraKle2012}. It is necessary to note, that the construction above is valid
under some genericity condition resulting in the cubic curve:
\begin{equation*}
x-tu=-u^s,\quad s=3
\end{equation*}
in dispersionless limit. The degenerate case with $s=2k+1$, $k\in\mathbb{N}$
was considered in~\cite{Kud1994, Cla2012} (see also~\cite{Kam2019}, where
the damb-breaking problem was investigated) leading to the ODEs of order $2k$
from Painlev\'{e} I hierarchy. These ODEs are nothing but a specific
combination of stationary reductions of higher (autonomous and
non-autonomous) symmetries of KdV equation~\cite{Kud1994, Adl2020}, which
can be obtained using Lax representation with additional linear PDE defining
evolution with respect to spectral parameter~\cite{Its1985, AdShaYa2000,
Kud2002}. Generalization on two-component systems can be found in~\cite%
{DubGrMol2015}, but for generic case only.

In the present work using the similar approach we present a new view to the
theory of isomonodromic deformations: the natural extension of the Tsarev
algebraic formula (\ref{etghm}). We restrict our consideration to such
well-known integrable dispersive systems like the KdV equation, the
Kaup--Boussinesq system in two-dimensional case. In three-dimensional case
we consider the Mikhal\"{e}v equation.

%Our \textbf{Motivation} is: the famous Gurevich--Pitaevskii problem (GP) is connected with the isomonodromic solution, written in the above form%
%\begin{equation*}
%x+ut=....????,
%\end{equation*}%
%which has a dispersionless limit%
%\begin{equation*}
%x+ut=u^{3}.
%\end{equation*}%
%This is a particular solution of the dispersionless KdV equation (i.e. the Hopf equation) $u_{t}=uu_{x}$. Nowadays the theory of dispersive shock waves (DSW) is widely known and widely applicable. It is well known that usually arbitrary solutions of quasilinear systems of first order are local only. This means they exist just a short time, while integrable dispersive systems possess infinitely many global solutions (i.e. multi-phase solutions). Thus, the GP problem allows to investigated the so called gradient catastrophe (or undular bores) for different physical phenomenons described by integrable dispersive systems.

\section{Energy Dependent Schr\"{o}dinger Case}

The Lax pair (here $\psi (x,\lambda )$ solves two equations simultaneously,
while the functions $U(x,\lambda )$ and $b(x,\lambda )$ are unknown, but not
arbitrary)%
\begin{equation}
\epsilon ^{2}\psi _{xx}=U\psi ,\text{ \ }\psi _{\lambda }=b\psi _{x}-\frac{1%
}{2}b_{x}\psi  \label{Lax1}
\end{equation}%
determines%
\begin{equation}
U_{\lambda }=\left( -\frac{1}{2}\epsilon ^{2}\partial _{x}^{3}+2U\partial
_{x}+U_{x}\right) b.  \label{comp1}
\end{equation}%
Two unknown functions and the single equation.
However, due to well-known results associated with this equation for the
case, where the independent variable $t$ is presented instead of the
spectral parameter $\lambda $, we are able to extract \textit{infinitely
many ordinary differential equations equipped by the above Lax representation%
}.
Indeed, the simplest substitution%
\begin{equation}
U=\lambda +u_{1}(x)  \label{kdv}
\end{equation}%
leads to%
\begin{equation*}
1=-\frac{1}{2}\epsilon ^{2}b_{xxx}+2(\lambda +u_{1})b_{x}+bu_{1,x}.
\end{equation*}%
The function $b(x,\lambda )$ has infinitely many particular monic polynomial
solutions. The simplest ansatz%
\begin{equation}
b=\lambda +b_{1}(x)  \label{mix1}
\end{equation}%
implies%
\begin{equation*}
1=\frac{1}{4}\epsilon ^{2}u_{1,xxx}-\frac{3}{2}u_{1}u_{1,x},\quad b_{1}=-%
\frac{1}{2}u_{1}.
\end{equation*}%
The first equation can be integrated once%
\begin{equation}
\epsilon ^{2}u_{1,xx}=3u_{1}^{2}+4x.  \label{P1}
\end{equation}%
This is the Painlev\'{e} I equation.

The second choice%
\begin{equation}
U=\lambda ^{2}+\lambda u_{1}(x)+u_{2}(x)  \label{kb}
\end{equation}%
leads%
\begin{equation*}
2\lambda +u_{1}=-\frac{1}{2}\epsilon ^{2}b_{xxx}+2(\lambda ^{2}+\lambda
u_{1}+u_{2})b_{x}+\lambda bu_{1,x}+bu_{2,x}.
\end{equation*}%
The function $b(x,\lambda )$ has infinitely many particular monic polynomial
solutions. The simplest ansatz (cf. (\ref{mix1}))%
\begin{equation*}
b=\lambda +b_{1}(x)
\end{equation*}%
implies%
\begin{equation*}
2=u_{2,x}-\frac{3}{2}u_{1}u_{1,x},\text{ \ \ }u_{1}=\frac{1}{4}\epsilon
^{2}u_{1,xxx}-u_{2}u_{1,x}-\frac{1}{2}u_{1}u_{2,x}, \quad b_{1}=-\frac{1}{2}%
u_{1}.
\end{equation*}%
First two equations can be integrated once, then we obtain the Painlev\'{e}
II equation%
\begin{equation}
\epsilon ^{2}u_{1,xx}=2u_{1}^{3}+8xu_{1}+\alpha,  \label{P2}
\end{equation}%
where%
\begin{equation*}
u_{2}=2x+\frac{3}{4}u_{1}^{2}.
\end{equation*}

Below we utilize these Painlev\'{e} equations (as well as their higher
analogues) for presentation of corresponding particular solutions of the
Korteweg--de Vries equation (KdV) and the Kaup--Boussinesq system (KB) in an
implicit form well known in the theory of semi-Hamiltonian hydrodynamic type
systems as the Tsarev Generalised Hodograph Method.

\section{2D Integrable Systems and Lax Triads}

The compatibility condition of the scalar Lax pair (cf. (\ref{Lax1}); here $%
\psi (x,\lambda )$ solves two equations simultaneously, while the functions $%
U(x,t,\lambda )$ and $a(x,t,\lambda )$ are unknown, but not arbitrary)%
\begin{equation}
\epsilon ^{2}\psi _{xx}=U\psi ,\text{ \ }\psi _{t}=a\psi _{x}-\frac{1}{2}%
a_{x}\psi  \label{Lax2}
\end{equation}%
yields (cf. (\ref{comp1}))%
\begin{equation}
U_{t}=\left( -\frac{1}{2}\epsilon ^{2}\partial _{x}^{3}+2U\partial
_{x}+U_{x}\right) a.  \label{comp2}
\end{equation}%
Two unknown functions and the single equation.

This equation contains infinitely many monic polynomial solutions with
respect to a spectral parameter $\lambda $, where%
\begin{equation}
U=\lambda ^{M}+\lambda ^{M-1}u_{1}(x,t)+\lambda
^{M-2}u_{2}(x,t)+...+u_{M}(x,t),  \label{monicu}
\end{equation}%
\begin{equation}
a=\lambda ^{K}+\lambda ^{K-1}a_{1}(x,t)+\lambda
^{K-2}a_{2}(x,t)+...+a_{K}(x,t).  \label{monica}
\end{equation}%
The Korteweg--de Vries equation is determined by the choice $M=K=1$, while
the Kaup--Boussinesq system is determined by the choice $M=2$, $K=1$.

Now we consider the scalar Lax triad instead of above Lax pairs (cf. (\ref%
{Lax1})), i.e.%
\begin{equation}
\epsilon ^{2}\psi _{xx}=U\psi ,\text{ \ }\psi _{t}=a\psi _{x}-\frac{1}{2}%
a_{x}\psi ,\text{ \ }\psi _{\lambda }=b\psi _{x}-\frac{1}{2}b_{x}\psi .
\label{Laxtriad}
\end{equation}%
The compatibility conditions lead to three consequences (cf. (\ref{comp1}), (%
\ref{comp2}))%
\begin{equation}
U_{t}=\left( -\frac{1}{2}\epsilon ^{2}\partial _{x}^{3}+2U\partial
_{x}+U_{x}\right) a,\text{ \ }U_{\lambda }=\left( -\frac{1}{2}\epsilon
^{2}\partial _{x}^{3}+2U\partial _{x}+U_{x}\right) b,\text{ \ }a_{\lambda
}+ab_{x}=b_{t}+ba_{x}.  \label{triadlax}
\end{equation}%
These three equations contain infinitely many monic polynomial solutions,
determined by (\ref{monicu}), (\ref{monica}) and%
\begin{equation}
b=\lambda ^{N}+\lambda ^{N-1}b_{1}(x,t)+\lambda
^{N-2}b_{2}(x,t)+...+b_{N}(x,t).  \label{nonlinearity}
\end{equation}
Consider the simplest case selected by the choice $M=N=K=1$:
\begin{equation*}
U=\lambda +u_{1}(x,t),\text{ \ }a=\lambda +a_{1}(x,t),\text{ \ }b=\lambda
+b_{1}(x,t).
\end{equation*}%
Then we obtain%
\begin{equation*}
u_{1,x}+2a_{1,x}=0,\text{ \ }u_{1,x}+2b_{1,x}=0,\text{ \ }b_{1,x}=a_{1,x},
\end{equation*}%
\begin{equation*}
u_{1,t}=-\frac{1}{2}\epsilon ^{2}a_{1,xxx}+2u_{1}a_{1,x}+a_{1}u_{1,x},\text{
\ }1=-\frac{1}{2}\epsilon ^{2}b_{1,xxx}+2u_{1}b_{1,x}+b_{1}u_{1,x},\text{ \ }%
1+a_{1}b_{1,x}=b_{1,t}+b_{1}a_{1,x}.
\end{equation*}%
Without loss of generality one can fix%
\begin{equation*}
a_{1}=-\frac{1}{2}u_{1},\quad b_{1}=\beta (t)-\frac{1}{2}u_{1},
\end{equation*}%
where the function $\beta (t)$ is not yet determined. Then three other
equations lead to the Korteweg--de Vries equation%
\begin{equation}
u_{1,t}=\frac{1}{4}\epsilon ^{2}u_{1,xxx}-\frac{3}{2}u_{1}u_{1,x},
\label{KdV}
\end{equation}%
together with two constraints%
\begin{equation*}
1-\beta (t)u_{1,x}=\frac{1}{4}\epsilon ^{2}u_{1,xxx}-\frac{3}{2}u_{1}u_{1,x},%
\text{ \ }3=2\beta ^{\prime }(t).
\end{equation*}%
Then we obtain%
\begin{equation*}
u_{1,t}=1-\frac{3}{2}tu_{1,x},\text{ \ }1-\frac{3}{2}tu_{1,x}=\frac{1}{4}%
\epsilon ^{2}u_{1,xxx}-\frac{3}{2}u_{1}u_{1,x}, \quad \beta (t)=\frac{3}{2}t.
\end{equation*}%
These relations are very important for understanding. The first equation is
linear and can be easily integrated:
\begin{equation}
u_{1}=p(z)+t,\text{ \ }z=x-\frac{3}{4}t^{2}.  \label{similarity}
\end{equation}%
The second equation is nonlinear, but ordinary differential equation, which
can be integrated once. Indeed ($\gamma (t)$ is an arbitrary function),%
\begin{equation}
x-\frac{3}{2}u_{1}t=\frac{1}{4}\epsilon ^{2}u_{1,xx}-\frac{3}{4}%
u_{1}^{2}+\gamma (t).  \label{kdvtsar}
\end{equation}%
However the consistency of the above equation with the KdV equation (\ref%
{KdV}) shows that function $\gamma (t)$ is constant, which can be eliminated
from the above equation by the shift of the independent variable $x$.

Under, the similarity reduction (\ref{similarity}), the above equation
becomes the Painlev\'{e} I equation%
\begin{equation*}
\epsilon ^{2}p^{\prime \prime }=3p^{2}+4z,
\end{equation*}%
which was already discussed in the previous section (cf. (\ref{P1})).

Nevertheless, here we would like emphasize that the ODE (\ref{kdvtsar}) has
exactly the form determined by the Tsarev Generalized Hodograph Method (see (%
\ref{etghm})). This means that the KdV equation (\ref{KdV}) has a particular
solution, determined by the Tsarev Generalized Hodograph Method (\ref%
{kdvtsar}). Simultaneously, this particular solution reduces to Painlev\'{e}
I by the similarity reduction (\ref{similarity}), while the KdV itself (\ref%
{KdV}) reduces to a derivative (with respect to the common independent
variable $z$) of the Painlev\'{e} I equation, i.e.%
\begin{equation*}
\epsilon ^{2}p^{\prime \prime \prime }=6pp^{\prime }+4.
\end{equation*}

Consider the second case selected by the choice $M=2$, $N=K=1$:
\begin{equation*}
U=\lambda ^{2}+\lambda u_{1}(x,t)+u_{2}(x,t),\text{ \ }a=\lambda +a_{1}(x,t),%
\text{ \ }b=\lambda +b_{1}(x,t).
\end{equation*}%
Then we obtain the Kaup--Boussinesq system%
\begin{equation}
u_{1,t}=u_{2,x}-\frac{3}{2}u_{1}u_{1,x},\text{ \ }u_{2,t}=\frac{1}{4}%
\epsilon ^{2}u_{1,xxx}-u_{2}u_{1,x}-\frac{1}{2}u_{1}u_{2,x},  \label{KB}
\end{equation}%
together with three constraints%
\begin{equation*}
2=(\beta (t)-\frac{3}{2}u_{1})u_{1,x}+u_{2,x},\text{ \ \ }u_{1}=\frac{1}{4}%
\epsilon ^{2}u_{1,xxx}-u_{2}u_{1,x}+(\beta (t)-\frac{1}{2}u_{1})u_{2,x},
\end{equation*}%
\begin{equation*}
\frac{1}{2}u_{1,t}+\frac{1}{2}\beta (t)u_{1,x}=\beta ^{\prime }(t)-1,
\end{equation*}%
Where without loss of generality one can fix%
\begin{equation*}
a_{1}=-\frac{1}{2}u_{1}, \quad b_{1}=\beta (t)-\frac{1}{2}u_{1},
\end{equation*}%
where the function $\beta (t)$ is not yet determined.

However, further compatibility conditions yield two linear equations%
\begin{equation*}
u_{1,t}+2tu_{1,x}=2,\text{ \ \ }u_{2,t}+2tu_{2,x}=u_{1},
\end{equation*}%
and two ordinary differential equations%
\begin{equation}
2=(2t-\frac{3}{2}u_{1})u_{1,x}+u_{2,x},\text{ \ \ }u_{1}=\frac{1}{4}\epsilon
^{2}u_{1,xxx}-u_{2}u_{1,x}+(2t-\frac{1}{2}u_{1})u_{2,x},  \label{pre}
\end{equation}%
where%
\begin{equation*}
\beta (t)=2t.
\end{equation*}

So the above linear equations have a general solution parameterized by two
arbitrary functions of a single variable%
\begin{equation}
u_{1}=p(z)+2t,\text{ \ }u_{2}=q(z)+tp(z)+t^{2},\text{ \ \ }z=x-t^{2}.
\label{similar}
\end{equation}%
Then above two ordinary differential equations take the form%
\begin{equation*}
2+\frac{3}{2}pp^{\prime }=q^{\prime },\text{ \ \ }p+p^{\prime }q=\frac{1}{4}%
\epsilon ^{2}p^{\prime \prime \prime }-\frac{1}{2}pq^{\prime }.
\end{equation*}%
The first equation can be integrated once, i.e.%
\begin{equation}
2z+\frac{3}{4}p^{2}=q,  \label{vw}
\end{equation}%
while the second becomes (we eliminated $q$)%
\begin{equation*}
2p+2zp^{\prime }+\frac{3}{2}v^{2}p^{\prime }=\frac{1}{4}\epsilon
^{2}p^{\prime \prime \prime }.
\end{equation*}%
This equation can be integrated once, i.e. ($\alpha $ is an arbitrary
constant)%
\begin{equation}
\epsilon ^{2}p^{\prime \prime }=2p^{3}+8zp+\alpha .  \label{p2}
\end{equation}%
Again this is Painlev\'{e} II equation (cf. (\ref{P2})).

However, we can integrate system (\ref{pre}) directly. Indeed, the first
equation becomes ($\delta (t)$ is an arbitrary function)%
\begin{equation}
x-u_{1}t=\frac{1}{2}u_{2}-\frac{3}{8}u_{1}^{2}+\delta (t),  \label{a}
\end{equation}%
while the second also can be integrated once ($\gamma (t)$ is an arbitrary
function):
\begin{equation}
xu_{1}=\frac{1}{4}\epsilon ^{2}u_{1,xx}-\frac{1}{8}u_{1}^{3}-\frac{1}{2}%
u_{1}u_{2}+2tu_{2}+\frac{1}{2}t u_{1}^{2}+\delta (t)u_{1}+\gamma (t).
\label{b}
\end{equation}%
However the consistency of the above equations with the Kaup--Boussinesq
system (\ref{KB}) shows that both functions $\gamma (t)$ and $\delta (t)$
are constants and $\delta $ can be eliminated by the shift of independent
variable $x$.

Introducing \textquotedblleft quasi-Riemann invariants\textquotedblright\ $%
r_{k}(x,t)$ such that
\begin{equation}
U=\lambda ^{2}+\lambda u_{1}(x,t)+u_{2}(x,t)=(\lambda -r_{1}(x,t))(\lambda
-r_{2}(x,t)),  \label{factor}
\end{equation}%
one can rewrite the Kaup--Boussinesq system in the symmetric form (see \cite%
{Pav2014})%
\begin{equation}
r_{1,t}=\frac{1}{2}(r_{2}+3r_{1})r_{1,x}+\frac{\epsilon ^{2}}{4}\frac{%
r_{1,xxx}+r_{2,xxx}}{r_{1}-r_{2}},\quad r_{2,t}=\frac{1}{2}%
(r_{1}+3r_{2})r_{2,x}-\frac{\epsilon ^{2}}{4}\frac{r_{1,xxx}+r_{2,xxx}}{%
r_{1}-r_{2}},  \label{1}
\end{equation}%
while both equations (\ref{a}) and (\ref{b}) simultaneously also can be
rewritten in the symmetric form%
\begin{equation}
x+\frac{1}{2}(r_{2}+3r_{1})t=-\frac{5}{8}r_{1}^{2}-\frac{1}{4}r_{1}r_{2}-%
\frac{1}{8}r_{2}^{2}-\frac{1}{4}\frac{\epsilon
^{2}(r_{1,xx}+r_{2,xx})+4\gamma }{r_{1}-r_{2}},  \label{2}
\end{equation}%
\begin{equation}
x+\frac{1}{2}(r_{1}+3r_{2})t=-\frac{1}{8}r_{1}^{2}-\frac{1}{4}r_{1}r_{2}-%
\frac{5}{8}r_{2}^{2}+\frac{1}{4}\frac{\epsilon
^{2}(r_{1,xx}+r_{2,xx})+4\gamma }{r_{1}-r_{2}}.  \label{3}
\end{equation}

\textbf{Remark}: Under the substitution%
\begin{equation*}
\psi =\exp \left( \frac{1}{\epsilon }\int \varphi dx\right)
\end{equation*}%
Lax triad (\ref{Laxtriad}) takes the form%
\begin{equation*}
\varphi^{2}+\epsilon \varphi_{x}=U,\text{ \ }\varphi_{t}=\left( a\varphi-%
\frac{1}{2}\epsilon a_{x}\right) _{x},\text{ \ }\varphi_{\lambda }=\left(
b\varphi-\frac{1}{2}\epsilon b_{x}\right) _{x}.
\end{equation*}%
So, the dispersionless limit ($\epsilon =0$) leads to three equations%
\begin{equation*}
\varphi^{2}=U,\text{ \ }\varphi_{t}=(a\varphi)_{x},\text{ \ }%
\varphi_{\lambda }=(b\varphi)_{x},
\end{equation*}%
whose compatibility conditions are (see (\ref{triadlax}))%
\begin{equation*}
U_{t}=(2U\partial _{x}+U_{x})a,\text{ \ }U_{\lambda }=(2U\partial
_{x}+U_{x})b,\text{ \ }a_{\lambda }+ab_{x}=b_{t}+ba_{x}.
\end{equation*}

In the dispersionless limit the Kaup--Boussinesq system reduces to the
shallow water system%
\begin{equation*}
r_{1,t}=\frac{1}{2}(r_{2}+3r_{1})r_{1,x},\quad r_{2,t}=\frac{1}{2}%
(r_{1}+3r_{2})r_{2,x},
\end{equation*}%
together with its particular solution determined by classical algebraic
formula from the Tsarev Generalised Hodograph Method:%
\begin{equation*}
x+\frac{1}{2}(r_{2}+3r_{1})t=-\frac{5}{8}r_{1}^{2}-\frac{1}{4}r_{1}r_{2}-%
\frac{1}{8}r_{2}^{2}-\frac{\gamma }{r_{1}-r_{2}},
\end{equation*}%
\begin{equation*}
x+\frac{1}{2}(r_{1}+3r_{2})t=-\frac{1}{8}r_{1}^{2}-\frac{1}{4}r_{1}r_{2}-%
\frac{5}{8}r_{2}^{2}+\frac{\gamma }{r_{1}-r_{2}}.
\end{equation*}%
Thus, the above construction (\ref{1}), (\ref{2}), (\ref{3}) describes
\textit{the natural Extension of the Tsarev Generalised Hodograph Method to
the Kaup--Boussinesq system, which possesses a dispersionless limit}.

\subsection{Higher Isomonodromic Deformations of Korteveg--de Vries equation}

The Korteveg--de Vries equation is a simplest scalar example in the theory
of integrable systems. By this reason, plenty of properties of
multi-component integrable systems are not recognizable in a scalar case.
Therefore, we present a few higher order computations for the KdV equation
for further comparison with two-component Kaup--Boussinesq system.

The second order nonlinearity for isomonodromic deformations%
\begin{equation*}
U=\lambda -r_{1}(x,t),\quad a=\lambda +a_{1}(x,t),\quad b=\lambda
^{2}+\lambda b_{1}(x,t)+b_{2}(x,t)
\end{equation*}%
lead to ODE of fourth order ($\beta_{1}$ is an arbitrary constant)%
\begin{equation}
x+\frac{3}{2}r_{1}t+\frac{3}{4}\beta_{1}r_{1}^{2}+\frac{5}{8}r_{1}^{3}+\frac{%
5}{16}\epsilon ^{2}r_{1,x}^{2}+\frac{1}{8}\epsilon
^{2}(2\beta_{1}+5r_{1})r_{1,xx}+\frac{1}{16}\epsilon ^{4}r_{1,xxxx}=0,
\label{fourthkdv}
\end{equation}%
where%
\begin{equation*}
a_{1}=\frac{1}{2}r_{1},\text{ \ \ }b_{1}=\frac{1}{2}r_{1}+\beta_{1},\text{ \
}b_{2}=\frac{1}{8}\epsilon ^{2}r_{1,xx}+\frac{3}{8}r_{1}^{2}+\frac{1}{2}%
\beta_{1}r_{1}+\frac{3}{2}t,
\end{equation*}%
while the Korteweg--de Vries equation is (in comparison with previous
section $r_{1}=-u_{1}$)%
\begin{equation*}
r_{1,t}=\frac{3}{2}r_{1}r_{1,x}+\frac{1}{4}\epsilon ^{2}r_{1,xxx}.
\end{equation*}

\textbf{Remark}: (\ref{fourthkdv}) can be rewritten in the form%
\begin{equation*}
x+\frac{3}{2}r_{1}t+\frac{1}{4}\beta_{1}(3r_{1}^{2}+\epsilon ^{2}r_{1,xx})+%
\frac{5}{8}r_{1}^{3}+\frac{5}{16}\epsilon ^{2}(2r_{1,x}^{2}+2r_{1}r_{1,xx})+%
\frac{1}{16}\epsilon ^{4}r_{1,xxxx}=0,
\end{equation*}%
where first three terms determine the previous case (cf. (\ref{kdvtsar})).
This means, that the constant $\beta_{1}$ shows that the higher
nonlinearities automatically include lower nonlinearities.

The above equation in the dispersionless limit yields%
\begin{equation*}
x+\frac{3}{2}r_{1}t+\frac{3}{4}\beta_{1}r_{1}^{2}+\frac{5}{8}r_{1}^{3}=0.
\end{equation*}%
This is a most known and investigated example of isomonodromic deformation
in the theory of DSW (dispersive shock waves), connected with remarkable
Gurevich--Pitaevski Problem.

Obviously, all other higher isomonodromic deformations for the KDV have
similar structure~\cite{Kud1994,Adl2020}
\begin{equation*}
x+\frac{3}{2}r_{1}t=F_{N}(r_{1},\epsilon r_{1,x},\epsilon ^{2}r_{1,xx},...),
\end{equation*}%
where $F_{N}$ are a differential polynomials, whose dispersionless limit is $%
C_{0,N}(r_{1})^{N}+C_{1,N}(r_{1})^{N-1}+...+C_{N,N}$. Here $C_{k,m}$ are
constants.

\subsection{Higher Isomonodromic Deformations of Kaup--Boussinesq System}

Similar computations for the second order nonlinearity of isomonodromic
deformations for the Kaup--Boussinesq system yield the ODE system (here $%
\beta_{1}$ and $\gamma_{1}$ are arbitrary constants):%
\begin{align*}
&2x-2tu_1+\beta_{1}\left( \frac{3}{4}u_{1}^{2}-u_{2}\right) -\frac{5}{8}%
u_{1}^{3}+\frac{3}{2}u_{1}u_{2}-\frac{1}{4}\epsilon ^{2}u_{1,xx}=0, \\
&\gamma_{1}+2t\left( \frac{1}{4}u_{1}^{2}-u_{2}\right) -\beta_{1}\left(
\frac{1}{4}u_{1}^{3}-u_{1}u_{2}+\frac{1}{4}\epsilon ^{2}u_{1,xx}\right) +%
\frac{15}{64}u_{1}^{4}-\frac{9}{8}u_{1}^{2}u_{2}+\frac{3}{4}u_{2}^{2} \\
&\quad\qquad\qquad\qquad \qquad\qquad\qquad\qquad\qquad +\epsilon ^{2}\left(
\frac{5}{16}u_{1,x}^{2}+\frac{1}{2}u_{1}u_{1,xx}-\frac{1}{4}u_{2,xx}\right)
=0,
\end{align*}%
where%
\begin{equation*}
U=\lambda ^{2}+\lambda u_{1}+u_{2},\quad a=\lambda +a_{1},\quad b=\lambda
^{2}+\lambda b_{1}+b_{2}
\end{equation*}%
and%
\begin{equation*}
a_{1}=-\frac{1}{2}u_{1},\text{ \ }b_{1}=-\frac{1}{2}u_{1}+\beta_{1},\text{ \
}b_{2}=\frac{3}{8}u_{1}^{2}-\frac{1}{2}u_{2}-\frac{1}{2}\beta_{1}u_{1}+2t.
\end{equation*}%
Taking into account (see (\ref{factor})) $r_{1}+r_{2}=-u_{1}$ and $%
r_{1}r_{2}=u_{2}$, the above ODE system becomes%
\begin{align*}
x+\frac{1}{2}(r_{2}&+3r_{1})t=\frac{1}{64}%
(-15r_{1}^{2}r_{2}-35r_{1}^{3}-9r_{1}r_{2}^{2}-5r_{2}^{3})-\frac{\beta_{1}}{8%
}(5r_{1}^{2}+2r_{1}r_{2}+r_{2}^{2})-\frac{\gamma_1}{r_{1}-r_{2}} \\[1.0mm]
&-\frac{\epsilon ^{2}\beta_{1}}{4}\frac{r_{1,xx}+r_{2,xx} }{r_{1}-r_{2}}-%
\frac{\epsilon ^{2}}{16}\frac{%
2(5r_{1}+r_{2})r_{1,xx}+5r_{1,x}^{2}+2r_{1,x}r_{2,x}+5r_{2,x}^{2}+6(r_{1}+r_{2})r_{2,xx}%
}{r_{1}-r_{2}}, \\[2.0mm]
x+\frac{1}{2}(r_{1}&+3r_{2})t=\frac{1}{64}%
(-9r_{1}^{2}r_{2}-5r_{1}^{3}-15r_{1}r_{2}^{2}-35r_{2}^{3})-\frac{\beta_{1}}{8%
}(r_{1}^{2}+2r_{1}r_{2}+5r_{2}^{2})+\frac{\gamma_1}{r_{1}-r_{2}} \\[1.0mm]
&+\frac{\epsilon ^{2}\beta_{1}}{4}\frac{r_{1,xx}+r_{2,xx}}{r_{1}-r_{2}} +%
\frac{\epsilon ^{2}}{16}\frac{%
6(r_{1}+r_{2})r_{1,xx}+5r_{1,x}^{2}+2r_{1,x}r_{2,x}+5r_{2,x}^{2}+2(r_{1}+5r_{2})r_{2,xx}%
}{r_{1}-r_{2}}.
\end{align*}
Since the constant $\beta_{1}$ can be eliminated by a shift, finally we
obtain%
\begin{align*}
x+\frac{1}{2}(r_{2}+3r_{1})t&=-\frac{1}{64}%
(35r_{1}^{3}+15r_{1}^{2}r_{2}+9r_{1}r_{2}^{2}+5r_{2}^{3})-\frac{\gamma_1}{%
r_{1}-r_{2}} \\
&-\frac{\epsilon ^{2}}{16}\frac{%
6(r_{1}+r_{2})r_{2,xx}+5r_{1,x}^{2}+2r_{1,x}r_{2,x}+5r_{2,x}^{2}+2(5r_{1}+r_{2})r_{1,xx}%
}{r_{1}-r_{2}}, \\
x+\frac{1}{2}(r_{1}+3r_{2})t&=-\frac{1}{64}%
(35r_{2}^{3}+15r_{1}r_{2}^{2}+9r_{1}^{2}r_{2}+5r_{1}^{3})+\frac{\gamma_1}{%
r_{1}-r_{2}} \\
&+\frac{\epsilon ^{2}}{16}\frac{%
6(r_{1}+r_{2})r_{1,xx}+5r_{1,x}^{2}+2r_{1,x}r_{2,x}+5r_{2,x}^{2}+2(r_{1}+5r_{2})r_{2,xx}%
}{r_{1}-r_{2}}.
\end{align*}%
Thus, in the dispersionless limit the shallow water system%
\begin{equation*}
r_{1,t}=\frac{1}{2}(r_{2}+3r_{1})r_{1,x},\quad r_{2,t}=\frac{1}{2}%
(r_{1}+3r_{2})r_{2,x},
\end{equation*}%
has a particular solution determined by classical algebraic formula from the
Tsarev Generalised Hodograph Method:%
\begin{equation*}
x+\frac{1}{2}(r_{2}+3r_{1})t=-\frac{1}{64}%
(35r_{1}^{3}+15r_{1}^{2}r_{2}+9r_{1}r_{2}^{2}+5r_{2}^{3})-\frac{\gamma_1}{%
r_{1}-r_{2}},
\end{equation*}%
\begin{equation*}
x+\frac{1}{2}(r_{1}+3r_{2})t=-\frac{1}{64}%
(35r_{2}^{3}+15r_{1}r_{2}^{2}+9r_{1}^{2}r_{2}+5r_{1}^{3})+\frac{\gamma_1}{%
r_{1}-r_{2}}.
\end{equation*}

All other higher nonlinearities (\ref{nonlinearity}) lead to higher order
ODEs with the above structure (\ref{etghm}).

\section{3D Integrable Systems and Lax Quads}

In the case of three-dimensional integrable systems for construction of
isomonodromic deformations one needs four linear commuting differential
operators. For instance, for the Energy dependent Schr\"{o}dinger equation,
the Lax quad is%
\begin{equation}
\epsilon ^{2}\psi _{xx}=U\psi ,\text{ \ }\psi _{t}=a\psi _{x}-\frac{1}{2}%
a_{x}\psi ,\text{ \ }\psi _{y}=\tilde{a}\psi _{x}-\frac{1}{2}\tilde{a}%
_{x}\psi ,\text{ \ }\psi _{\lambda }=b\psi _{x}-\frac{1}{2}b_{x}\psi .
\label{laxquad}
\end{equation}%
Corresponding compatibility conditions yield six differential consequences%
\begin{equation}
U_{t}=\left( -\frac{1}{2}\epsilon ^{2}\partial _{x}^{3}+2U\partial
_{x}+U_{x}\right) a,\text{ \ }\tilde{a}_{\lambda }+\tilde{a}b_{x}=b_{y}+b%
\tilde{a}_{x},  \label{6}
\end{equation}%
\begin{equation}
U_{y}=\left( -\frac{1}{2}\epsilon ^{2}\partial _{x}^{3}+2U\partial
_{x}+U_{x}\right) \tilde{a},\text{ \ }a_{\lambda }+ab_{x}=b_{t}+ba_{x},
\label{4}
\end{equation}%
\begin{equation}
U_{\lambda }=\left( -\frac{1}{2}\epsilon ^{2}\partial _{x}^{3}+2U\partial
_{x}+U_{x}\right) b,\text{ \ }a_{y}+a\tilde{a}_{x}=\tilde{a}_{t}+\tilde{a}%
a_{x}.  \label{5}
\end{equation}

This system possesses infinitely many monic polynomial solutions determined
by (all coefficients of these polynomials (\ref{monicu}), (\ref{monica}), (%
\ref{nonlinearity}) depend on $x,t$ and $y$)%
\begin{equation}
\tilde{a}=\lambda ^{L}+\lambda ^{L-1}a_{1}(x,t,y)+\lambda
^{L-2}a_{2}(x,t,y)+...+a_{L}(x,t,y).  \label{monicb}
\end{equation}

In this article without loss of generality we restrict our consideration to
the remarkable Mikhal\"{e}v 3D integrable system (see the last equation in (%
\ref{5}))%
\begin{equation*}
a_{2,x}=a_{1,t},\text{ \ }a_{1,y}+a_{1}a_{2,x}=a_{2,t}+a_{2}a_{1,x},
\end{equation*}%
determined by the simplest choice $a=\lambda +a_{1}(x,t)$ and $\tilde{a}%
=\lambda ^{2}+\lambda a_{1}(x,t)+a_{2}(x,t)$.

\subsection{Mikhal\"{e}v System and Two Commuting KdV Flows}

In this case, the particular solution of the Mikhal\"{e}v system is
determined by%
\begin{equation*}
x+\frac{3}{2}r_{1}t+\frac{5}{8}(3r_{1}^{2}+\epsilon ^{2}r_{1,xx})y+\frac{5}{8%
}r_{1}^{3}+\frac{5}{16}\epsilon ^{2}(r_{1,x}^{2}+2r_{1}r_{1,xx})+\frac{1}{16}%
\epsilon ^{4}r_{1,xxxx}=0,
\end{equation*}%
where%
\begin{equation*}
a_{1}=\frac{1}{2}r_{1},\quad a_{2}=\frac{1}{8}\epsilon ^{2}r_{1,xx}+\frac{3}{%
8}r_{1}^{2}.
\end{equation*}%
The corresponding Lax quad (\ref{laxquad}) is determined by the choice%
\begin{equation*}
U=\lambda -r_{1},\quad a=\lambda +a_1,\quad \tilde{a}=\lambda ^{2}+\lambda
a_{1}+a_{2},\text{ \ }b=\lambda ^{2}+\lambda \left( \frac{1}{2}r_{1}+\frac{5%
}{2}y\right) +\frac{1}{8}\epsilon ^{2}r_{1,xx}+\frac{3}{8}r_{1}^{2}+\frac{5}{%
4}yr_{1}+\frac{3}{2}t.
\end{equation*}%
Here evolution of $r_{1}(x,t,y)$ with respect to $x$ and $t$ is described by
the KdV equation
\begin{equation*}
r_{1,t}=\frac{3}{2}r_{1}r_{1,x}+\frac{1}{4}\epsilon ^{2}r_{1,xxx},
\end{equation*}%
while its evolution with respect to $x$ and $y$ satisfies to the higher
commuting flow from the KdV hierarchy%
\begin{equation*}
r_{1,y}=\frac{15}{8}r_{1}^{2}r_{1,x}+\frac{5}{8}\epsilon
^{2}(2r_{1,x}r_{1,xx}+r_{1}r_{1,xxx})+\frac{1}{16}\epsilon ^{4}r_{1,xxxxx}.
\end{equation*}

\subsection{Mikhal\"{e}v System and Two Commuting KB Flows}

In this case, the particular solution of the Mikhal\"{e}v system is
determined by a dispersive analogue of the Tsarev Generalized Hodograph
Method%
\begin{align*}
x+\frac{1}{2}t(r_{2}+3r_{1})+y\left( \frac{15}{8}r_{1}^{2}+\frac{3}{4}%
r_{1}r_{2}+\frac{3}{8}r_{2}^{2}+\frac{3\epsilon ^{2}}{4}\frac{%
r_{1,xx}+r_{2,xx}}{r_{1}-r_{2}}\right) =\frac{1}{64}\left(
-15r_{1}^{2}r_{2}-35r_{1}^{3}-9r_{1}r_{2}^{2}-5r_{2}^{3}\right) & \\[1mm]
-\frac{\gamma _{1}}{r_{1}-r_{2}}-\frac{\epsilon ^{2}}{16}\frac{%
2(5r_{1}+r_{2})r_{1,xx}+5r_{1,x}^{2}+2r_{1,x}r_{2,x}+5r_{2,x}^{2}+6(r_{1}+r_{2})r_{2,xx}%
}{r_{1}-r_{2}}& , \\[2mm]
x+\frac{1}{2}t(r_{1}+3r_{2})+y\left( \frac{3}{8}r_{1}^{2}+\frac{3}{4}%
r_{1}r_{2}+\frac{15}{8}r_{2}^{2}-\frac{3\epsilon ^{2}}{4}\frac{%
r_{1,xx}+r_{2,xx}}{r_{1}-r_{2}}\right) =\frac{1}{64}\left(
-9r_{1}^{2}r_{2}-5r_{1}^{3}-15r_{1}r_{2}^{2}-35r_{2}^{3}\right) & \\[1mm]
 +\frac{\gamma _{1}}{r_{1}-r_{2}}+\frac{\epsilon ^{2}}{16}\frac{%
6(r_{1}+r_{2})r_{1,xx}+5r_{1,x}^{2}+2r_{1,x}r_{2,x}+5r_{2,x}^{2}+2(r_{1}+5r_{2})r_{2,xx}%
}{r_{1}-r_{2}}&,
\end{align*}%
where%
\begin{equation*}
a_{1}=\frac{1}{2}(r_{1}+r_{2}),\quad a_{2}=\frac{1}{4}r_{1}r_{2}+\frac{3}{8}%
r_{1}^{2}+\frac{3}{8}r_{2}^{2}.
\end{equation*}%
The corresponding Lax quad (\ref{laxquad}) is determined by the choice
\begin{equation*}
U=\lambda ^{2}+\lambda u_{1}+u_{2},\quad a=\lambda +a_{1},\quad \tilde{a}%
=\lambda ^{2}+\lambda a_{1}+a_{2},\text{ \ }b=\lambda ^{2}+\lambda
b_{1}+b_{2},
\end{equation*}%
where.%
\begin{equation*}
u_{1}=-(r_{1}+r_{2}),\text{ \ \ }u_{2}=r_{1}r_{2},\text{ \ }b_{1}=\frac{1}{2}%
(r_{1}+r_{2})+3y,\quad b_{2}=2t+\frac{3}{2}y(r_{1}+r_{2})+\frac{1}{4}%
r_{1}r_{2}+\frac{3}{8}r_{1}^{2}+\frac{3}{8}r_{2}^{2}.
\end{equation*}%
Here evolution of $r_{k}(x,t,y)$ with respect to $x$ and $t$ is described by
the KB system%
\begin{equation*}
r_{1,t}=\frac{1}{2}(r_{2}+3r_{1})r_{1,x}+\frac{\epsilon ^{2}}{4}\frac{%
r_{1,xxx}+r_{2,xxx}}{r_{1}-r_{2}},\quad r_{2,t}=\frac{1}{2}%
(r_{1}+3r_{2})r_{2,x}-\frac{\epsilon ^{2}}{4}\frac{r_{1,xxx}+r_{2,xxx}}{%
r_{1}-r_{2}},
\end{equation*}%
while their evolution with respect to $x$ and $y$ satisfies to the higher
commuting flow from the KB hierarchy
\begin{align*}
r_{1,y}=\left( \frac{15}{8}r_{1}^{2}+\frac{3}{4}r_{1}r_{2}+\frac{3}{8}%
r_{2}^{2}\right) r_{1,x}+\frac{\epsilon ^{2}}{8}\frac{%
3(3r_{1,x}+r_{2,x})r_{1,xx}+3\left( r_{1,x}+3r_{2,x}\right) r_{2,xx}}{%
r_{1}-r_{2}}& \\
+\frac{\epsilon ^{2}}{8}\frac{3\left( r_{1}+r_{2}\right) r_{2,xxx}+\left(
5r_{1}+r_{2}\right) r_{1,xxx}}{r_{1}-r_{2}}& , \\
r_{2,y}=\left( \frac{3}{8}r_{1}^{2}+\frac{3}{4}r_{1}r_{2}+\frac{15}{8}%
r_{2}^{2}\right) r_{2,x}-\frac{\epsilon ^{2}}{8}\frac{%
3(3r_{1,x}+r_{2,x})r_{1,xx}+3(r_{1,x}+3r_{2,x})r_{2,xx}}{r_{1}-r_{2}}& \\
-\frac{\epsilon ^{2}}{8}\frac{3\left( r_{1}+r_{2}\right) r_{1,xxx}+\left(
r_{1}+5r_{2}\right) r_{2,xxx}}{r_{1}-r_{2}}& .
\end{align*}

\section{Conclusion}

In this paper we demonstrated that integrable (by the Inverse Scattering
Transform) dispersive systems (possessing a dispersionless limit) have a
natural class of solutions (determined by ODEs known as isomonodromic
deformations), which can be written exactly as a differential extension of
the Tsarev Algebraic Formula%
\begin{equation}
x+tv^{i}(\mathbf{r,}\epsilon \mathbf{r}_{x}\mathbf{,}\epsilon ^{2}\mathbf{r}%
_{xx},...)=w^{i}(\mathbf{r,}\epsilon \mathbf{r}_{x}\mathbf{,}\epsilon ^{2}%
\mathbf{r}_{xx},...),  \label{godo}
\end{equation}%
whose dispersionless limit ($\epsilon =0$) is given by%
\begin{equation*}
x+tv_{0}^{i}(\mathbf{r})=w_{0}^{i}(\mathbf{r}),
\end{equation*}%
where%
\begin{equation*}
v^{i}(\mathbf{r,}\epsilon \mathbf{r}_{x}\mathbf{,}\epsilon ^{2}\mathbf{r}%
_{xx},...)=v_{0}^{i}(\mathbf{r})+\epsilon v_{1s}^{i}(\mathbf{r}%
)r_{s,x}+\epsilon ^{2}(v_{2s}^{i}(\mathbf{r})r_{s,xx}+v_{2ks}^{i}(\mathbf{r}%
)r_{k,x}r_{s,x})+...
\end{equation*}%
\begin{equation*}
w^{i}(\mathbf{r,}\epsilon \mathbf{r}_{x}\mathbf{,}\epsilon ^{2}\mathbf{r}%
_{xx},...)=w_{0}^{i}(\mathbf{r})+\epsilon w_{1s}^{i}(\mathbf{r}%
)r_{s,x}+\epsilon ^{2}(w_{2s}^{i}(\mathbf{r})r_{s,xx}+w_{2ks}^{i}(\mathbf{r}%
)r_{k,x}r_{s,x})+...
\end{equation*}

Here we considered the two-component Kaup--Boussinesq system (as a simple
nontrivial example), determined by the Lax pair%
\begin{equation*}
\epsilon ^{2}\psi _{xx}=(\lambda ^{2}+\lambda u_{1}+u_{2})\psi ,\text{ \ }%
\psi _{t}=(\lambda -\frac{1}{2}u_{1})\psi _{x}+\frac{1}{4}u_{1,x}\psi .
\end{equation*}%
The Kaup--Boussinesq system possesses infinitely many particular solutions
(isomonodromic deformations) selected by the third linear equation (here all
these three linear equations we call the Lax triad)%
\begin{equation*}
\psi _{\lambda }=b\psi _{x}-\frac{1}{2}b_{x}\psi ,
\end{equation*}%
where%
\begin{equation*}
b=\lambda ^{N}+\lambda ^{N-1}b_{1}(x,t)+\lambda
^{N-2}b_{2}(x,t)+...+b_{N}(x,t).
\end{equation*}%
The natural number $N$ selects different degrees of nonlinearity, which
describe corresponding DSW scenario. So all of them can be written in the
form (\ref{godo}).

We showed that commuting flows in~\eqref{godo} in two-component case
generated by the method of isomonodromic deformations are essentially not
self-similar, while every particular higher commuting flow is homogeneous~%
\cite{Kri1988,Kri1989,Pot1988}.

Obviously, the presented construction can be generalised to a
multi-dimensional case (for instance, the
Kadomtsev--Petviashvili equation) and to a multi-component case (for
instance, higher order degrees of the Energy dependent Schr\"{o}dinger
equation and Gelfand--Dickey reductions of the Kadomtsev--Petviashvili
hierarchy).

Three-component case is a first interesting example in the Topological Field
Theory \cite{Dub, Krichever1, Krichever2}, which needs significantly more computations,
will be considered in the next paper.

\section{Acknowledgements}

ZVM gratefully acknowledges the support from National Natural Science
Foundation of China (Grant No.~12471239) and the Guangdong Basic and Applied
Basic Research Foundation (Grant No.~2024A1515013106). MVP's work was
partially supported by the NSFC-RFBR (grant 12111530003), by the SAFEA
(grant G2023018006L) and by the NSFC (grant 12431008).

MVP also thanks V.E. Adler, A.V. Aksenov, E.V. Ferapontov, G.A. El, B.I.
Suleimanov for important discussions.


\addcontentsline{toc}{section}{References}

\begin{thebibliography}{99}
\bibitem{Ts1985} Tsarev, S.P., 1985. On Poisson brackets and one-dimensional
systems of hydrodynamic type. In Sov. Math. Doklady (Vol. 31, p. 488).

\bibitem{Ts1991} Tsarev, S.P., 1991. The geometry of Hamiltonian systems of
hydrodynamic type. The generalized hodograph method. Mathematics of the
USSR-Izvestiya, 37(2), p.397.

\bibitem{Dub} Dubrovin, B.A., Geometry of 2D topological field theories,
Lecture Notes in Mathematics, V.1620, Berlin, Springer, 120-348.


\bibitem{Krichever1} Krichever, I.M. \newblock The dispersionless equations
and topological minimal models, Comm. Math. Phys., \textbf{143, }No. 2
(1992) 415-429.

\bibitem{Krichever2} Krichever, I.M. \newblock The $\tau $-function of the
universal Whitham hierarchy, matrix models and topological field theories,
Comm. Pure Appl. Math. \textbf{47} (1994) 437-475.


\bibitem{VeSha1993} Veselov, A.P. and Shabat, A.B., 1993. Dressing chains
and the spectral theory of the Schr\"{o}dinger operator. Functional Analysis
and Its Applications, 27(2), pp.81-96.

\bibitem{GP1973} Gurevich, A.V. and Pitaevskii, L.P., 1973. Nonstationary
structure of a collisionless shock wave. Zhurnal Eksperimentalnoi i
Teoreticheskoi Fiziki, 65, pp.590-604.

\bibitem{Sul1994} Suleimanov, B.I., 1994. Onset of nondissipative shock
waves and the nonperturbative quantum theory of gravitation. Zhurn. Eskper.
Teor. Fiz, 105(5), pp.1089-1099.

\bibitem{KudSul1996} Kudashev, V. and Suleimanov, B., 1996. A soft mechanism
for the generation of dissipationless shock waves. Physics Letters A,
221(3-4), pp.204-208.

\bibitem{Dubr2006} Dubrovin, B.A., 2006. On Hamiltonian perturbations of
hyperbolic systems of conservation laws, II: universality of critical
behaviour. Communications in mathematical physics, 267, pp.117-139.


\bibitem{ClaGra2009} Claeys, T. and Grava, T., 2009. Universality of the
break-up profile for the KdV equation in the small dispersion limit using
the Riemann--Hilbert approach. Communications in mathematical physics,
286(3), pp.979-1009.

\bibitem{GraKle2012} Grava, T. and Klein, C., 2012. A numerical study of the
small dispersion limit of the Korteweg--de Vries equation and
asymptotic solutions. Physica D: Nonlinear Phenomena, 241(23-24),
pp.2246-2264.

\bibitem{Kud1994} Kudashev, V.R., 1994. KdV shock-like waves as invariant
solutions of KdV equation symmetries. arXiv preprint patt-sol/9404002.

\bibitem{Cla2012} Claeys, T., 2012. Pole-free solutions of the first Painlev%
\'{e}\ hierarchy and non-generic critical behavior for the KdV equation.
Physica D: Nonlinear Phenomena, 241(23-24), pp.2226-2236.

\bibitem{Kam2019} Kamchatnov, A.M., 2019. Self-similar wave breaking in
dispersive Korteweg--de Vries hydrodynamics. Chaos: An Interdisciplinary
Journal of Nonlinear Science, 29(2).

\bibitem{Adl2020} Adler, V.E., 2020. Nonautonomous symmetries of the KdV
equation and step-like solutions. Journal of Nonlinear Mathematical Physics,
27(3), pp.478-493.

\bibitem{Its1985} Its, A.R., 1985. Isomonodromic solutions of the zero
curvature equations. Izv. Akad. Nauk SSSR Ser. Mat, 49, pp.530-565.

\bibitem{AdShaYa2000} Adler, V.E., Shabat, A.B. and Yamilov, R.I., 2000.
Symmetry approach to the integrability problem. Theoretical and Mathematical
Physics, 125(3), pp.1603-1661.

\bibitem{Kud2002} Kudryashov, N.A., 2002. Fourth-order analogies to the
Painlev\'{e}\ equations. Journal of Physics A: Mathematical and General,
35(21), p.4617.

\bibitem{Pav2014} Pavlov, M.V., 2014. Integrable dispersive chains and
energy dependent Schr\"{o}dinger operator. Journal of Physics A:
Mathematical and Theoretical, 47(29), p.295204.

\bibitem{DubGrMol2015} Dubrovin, B., Grava, T., Klein, C. and Moro, A.,
2015. On critical behaviour in systems of Hamiltonian partial differential
equations. Journal of Nonlinear Science, 25(3), pp.631-707.

\bibitem{Kri1988} Krichever, I.M., 1988. Method of averaging for
two-dimensional ``integrable'' equations. Functional analysis and its
applications, 22(3), pp.200-213.

\bibitem{Kri1989} Krichever, I.M., 1989. Spectral theory of two-dimensional
periodic operators and its applications. Russian mathematical surveys,
44(2), p.145.

\bibitem{Pot1988} Potemin, G.V., 1988. Algebro-geometric construction of
self-similar solutions of the Whitham equations. Russian Mathematical
Surveys, 43(5), p.252.
\end{thebibliography}
\end{document}